

\input harvmac

\overfullrule=0pt


\def\A{{\scriptscriptstyle A}}

\def\L{{\scriptscriptstyle L}}

\def\R{{\scriptscriptstyle R}}


\def\CA{{\cal A}}

\def\CL{{\cal L}}


\def\aEM{\alpha_{\scriptscriptstyle EM}}

\def\bar#1{\overline{#1}}
\def\Br{{\rm Br}}
\def\dash{{\> \over \,\>}} 		

\def\GeV{\>\, \rm GeV}

\def\gtap{\raise.3ex\hbox{$>$\kern-.75em\lower1ex\hbox{$\sim$}}}
\def\ltap{\raise.3ex\hbox{$<$\kern-.75em\lower1ex\hbox{$\sim$}}}

\def\MeV{\,\> {\rm MeV}}
\def\mpi{m_{\pi}}

\def\mq{m_q}

\def\vslash{\rlap{/}v}


\def\b{\beta}

\def\g{\gamma}



\nref\WiseI{M. Wise, Phys. Rev. {\bf D45} (1992) R2188.}
\nref\Burdman{G. Burdman and J. Donoghue, Phys. Lett. {\bf B280} (1992) 287.}
\nref\Yan{H.Y. Cheng {\it et al.}, Phys. Rev. {\bf D46} (1992) 1148.}
\nref\ChoGeorgi{P. Cho and H. Georgi, Phys. Lett. {\bf B296} (1992) 408;
  (E) Phys. Lett. {\bf B300} (1993) 410.}
\nref\Amundson{J.F. Amundson {\it et al.}, Phys. Lett. {\bf B296} (1992) 415.}
\nref\YanII{H.Y. Cheng {\it et al.}, Phys. Rev. {\bf D47} (1992) 1030.}
\nref\CLEOI{F. Butler {\it et al.} (CLEO Collaboration), Phys. Rev. Lett.
 {\bf 69} (1992) 2041.}
\nref\GasserLeutwyler{J. Gasser and H. Leutwyler, Nucl. Phys. {\bf B250}
(1985) 465.}
\nref\CLEOII{M.S. Alam {\it et al.} (CLEO Collaboration), talk presented
 at the EPS meeting summer conference (1993).}
\nref\YanIII{H.Y. Cheng {\it et al.}, Cornell preprint CLNS 93/1189 (1993).}


\nfig\treegraph{Tree graph that mediates $D_s^* \to D_s \pi^0$. The solid
circle represents the isospin violating $\pi^0\dash\eta$ mixing vertex
proportional to $(m_d-m_u)$.}
\nfig\brplot{Isospin violating branching fraction $\Br(D_s^* \to D_s \pi^0)$
plotted against $\Br(D^{*+} \to D^+ \gamma)$.  $SU(3)$ corrections of order
$\mq^{1/2}$ to the light quark magnetic moments are included in this graph.}


\def\CITTitle#1#2#3{\nopagenumbers\abstractfont
\hsize=\hstitle\rightline{#1}
\vskip 0.6in\centerline{\titlefont #2} \centerline{\titlefont #3}
\abstractfont\vskip .5in\pageno=0}
\CITTitle{{\baselineskip=12pt plus 1pt minus 1pt
  \vbox{\hbox{CALT-68-1914}\hbox{DOE RESEARCH AND}\hbox{DEVELOPMENT
  REPORT}}}} {Comment on $D_s^* \to D_s \pi^0$ Decay}{}
\centerline{
  Peter Cho\footnote{$^\dagger$}{Work supported in part by an
  SSC Fellowship and by the U.S. Dept. of Energy under DOE Grant no.
  DE-FG03-92-ER40701.}
  and Mark B. Wise\footnote{$^\ddagger$}{Work supported in part
  by the U.S. Dept. of Energy under DOE Grant no. DE-FG03-92-ER40701.}}

\centerline{Lauritsen Laboratory}
\centerline{California Institute of Technology}
\centerline{Pasadena, CA  91125}

\vskip .3in
\centerline{\bf Abstract}
\bigskip

	We calculate the rate for $D_s^* \rightarrow D_s \pi^0$ decay using
Chiral Perturbation Theory.  This isospin violating process results from
$\pi^0\dash\eta$ mixing, and its amplitude is proportional to
$(m_d - m_u)/\bigl(m_s-(m_u+m_d)/2 \bigr)$.  Experimental information on the
branching ratio for $D_s^* \rightarrow D_s \pi^0$ can provide insight into
the pattern of $SU(3)$ violation in radiative $D^*$ decays.

\Date{1/94}


	The strong and radiative decays of $D^*$ mesons have been studied
in refs.~\refs{\WiseI{--}\YanII}\ using a synthesis of Chiral Perturbation
Theory and the Heavy Quark Effective Theory.  In this note, we extend
the analysis of such transitions to include the isospin violating mode
$D_s^* \rightarrow D_s \pi^0$.  We describe how experimental information on
this process can yield insight into the pattern of $SU(3)$ breaking in
radiative $D^*$ decays.

	Isospin violation enters into the low energy strong interactions of
the $\pi$, $K$ and $\eta$ pseudo-Goldstone bosons through their mass term
\eqn\massterm{\CL_{\rm mass} = {\mu f^2 \over 4} \Tr (\xi m_q \xi
+ \xi^{\dagger} m_q \xi^{\dagger})}
in the chiral Lagrange density.  Here $\xi=\exp(iM/f)$ represents a
$3 \times 3$ special unitary matrix that incorporates the meson octet
\eqn\Mmatrix{M = \left(\matrix{\pi^0/\sqrt{2} + \eta/\sqrt{6} & \pi^+ & K^+\cr
	\pi^- &  -\pi^0/\sqrt{2} + \eta/\sqrt{6} & K^0\cr
	K^- & \bar K^0 & - \sqrt{{2\over 3}} \eta\cr}\right),}
and $m_q$ denotes the light quark mass matrix
\eqn\massmatrix{m_q = \left(\matrix{m_u & 0 & 0\cr
	0 & m_d & 0\cr
	0 & 0 & m_s\cr}\right).}
The exponentiated Goldstone field and Lagrange density in \massterm\
transform as $\xi \to L\xi U^\dagger = U \xi R^\dagger$ and
$(3_L, \bar 3_R) + (\bar 3_L,3_R)$ under the chiral symmetry group
$SU(3)_\L \times SU(3)_\R$.  The Goldstone boson mass term contains the
off-diagonal interaction
\eqn\mixing{\CL_{\rm mixing} = \mu {(m_d - m_u)\over \sqrt{3}} \pi^0 \eta}
which mixes the $I=1$ neutral pion with the $I=0$ eta.  This isospin violating
mixing vanishes in the limit of equal up and down quark masses.

	The low energy interactions of pseudo-Goldstone bosons with mesons
containing a single heavy quark are described by the leading order chiral
Lagrange density
\eqn\Lagrangian{\eqalign{
\CL &= - i \Tr \bar H^a v_\mu \partial^\mu H_a + {i\over 2}
  \Tr \bar  H^a H_b v_\mu (\xi^{\dagger} \partial^\mu \xi + \xi\partial^\mu
  \xi^{\dagger})^b_a \cr
& \qquad + {ig\over 2} \Tr \bar H^a H_b \gamma_\mu \gamma_5 (\xi^{\dagger}
 \partial^\mu \xi - \xi \partial^\mu \xi^{\dagger})^b_a. \cr}}
The $4\times 4$ matrix
\eqn\Hfield{H_a = {(1 + \vslash)\over 2} (P_a^{*\mu} \gamma_\mu - P_a
\gamma_5)}
combines together velocity dependent pseudoscalar and vector meson fields
that carry a light quark flavor index $a$.  When the heavy quark
inside the meson is charm, the individual components of the fields in $H$
are $(D_1^{(*)},  D_2^{(*)}, D_3^{(*)}) = (D^{(*)0}, D^{(*)+}, D_s^{(*)})$.
$H$ transforms under heavy quark spin symmetry $SU(2)_v$ and chiral
$SU(3)_L \times SU(3)_R$ as $H_a \rightarrow S(HU^{\dagger})_a$ where
$S \in SU(2)_v$.

	The interaction term proportional to $g$ in Lagrange density
\Lagrangian\ mediates the strong decays $D^{*+} \rightarrow D^0 \pi^+$,
$D^{*+} \rightarrow  D^+ \pi^0$ and $D^{*0} \rightarrow D^0 \pi^0$.  The
tree level rate for the charged pion mode is
\eqn\strongrate{\Gamma (D^{*+} \rightarrow D^0 \pi^+) =
{g^2\over 6\pi f_\pi^2} |\vec p_{\pi^+}|^3,}
while the corresponding widths in the neutral pion channels are a factor of
two smaller due to isospin.  Unfortunately, the value for $g$ has not been
extracted from these single pion partial widths since the total widths
of charmed vector mesons are too narrow to be experimentally resolved.  The
coupling constant can therefore only be estimated in various models.  In the
nonrelativistic constituent quark model, one finds $g=1$,
\foot{This estimate is similar to the result $g_\A= 5/3$ for the pion nucleon
coupling.}
whereas $g=0.75$ in the chiral quark model.

	Charmed vector mesons can also decay to their pseudoscalar
counterparts via single photon emission.  The matrix elements for such
electromagnetic transitions have the general form
\eqn\amplitude{\CA (D_a^* \rightarrow D_a \gamma) = e\mu_a
 \epsilon^{\mu\nu\sigma\lambda} \varepsilon_\mu^* (\gamma)  v_\nu k_\sigma
 \varepsilon_\lambda (D^*)}
where $e\mu_a/2$ is the transition magnetic moment, $k$ is the photon
momentum, $\varepsilon(\gamma)$ is the photon polarization and
$\varepsilon(D^*)$ is the $D^*$ polarization.  After squaring the radiative
amplitude, averaging over the initial state polarization and summing over the
final state polarization, one finds the electromagnetic partial width
\eqn\photrate{\Gamma (D_a^* \rightarrow D_a \gamma) = {\aEM\over 3}
|\mu_a|^2 |\vec k|^3.}

	The transition magnetic moments which enter into the radiative matrix
element \amplitude\ receive contributions from photon couplings to both the
heavy charm and light quark electromagnetic currents
${2\over 3}\bar c\gamma_{\mu}c$ and
${2\over 3}\bar u\gamma_{\mu}u-{1\over 3}\bar d\gamma_{\mu}d-{1\over 3}\bar
s\gamma_{\mu}s$.  They consequently decompose as
$\mu_a = \mu^{(h)}+\mu_a^{(l)}$.  The charm magnetic moment is fixed by heavy
quark spin symmetry to be $\mu^{(h)} = 2/(3 m_c)$.  The $\mu_a^{(l)}$ moments
on the other hand are {\it a priori} undetermined.   In the $SU(3)$ symmetry
limit, they are proportional to the electric charges $Q_a$ of the light
quarks:
\eqn\lightpart{\mu_a^{(l)} = \beta Q_a.}
The proportionality constant $\beta$ has dimensions of inverse mass and
represents an unknown reduced matrix element.

	The strong and electromagnetic interactions compete in $D^*$ meson
decays.  The inherently weaker strength of the radiative modes is offset by
the limited phase space available in the strong interaction channels.  This
competition is unusual in the strange charmed sector where the small
$D_s^* \to D_s \gamma$ process dominates over the even smaller isospin
violating transition $D_s^* \to D_s \pi^0$.  The neutral pion mode proceeds at
tree level via virtual eta emission as illustrated in \treegraph.  The
intermediate $\eta$ converts into a $\pi^0$ through the mixing term in
eqn.~\mixing.  The $\eta$ propagator effectively renders the
$D_s^* \rightarrow D_s \pi^0$ amplitude inversely proportional to the strange
quark mass.  Consequently, the diagram is multiplied by the isospin
violation factor \GasserLeutwyler\
\eqn\quarkratio{(m_d - m_u) / \bigl(m_s - (m_u+m_d)/2 \bigr) \simeq 1/43.7}
which is larger than one might have naively guessed.  There is also
an electromagnetic contribution to the $D_s^* \to D_s \pi^0$ amplitude.  But
since it is down by $\aEM / \pi \simeq 1/430$, the electromagnetic term
is expected to be less important than the strong interaction contribution
which we focus upon here.

	A straightforward computation of the rate for the isospin violating
decay yields
\eqn\pizerorate{\Gamma (D_s^* \rightarrow D_s \pi^0) =
{g^2\over 48\pi f_{\eta}^2} \left({m_d - m_u\over m_s-(m_u+m_d)/2}\right)^2
|\vec p_ {\pi^0} |^3.}
This partial width sensitively depends upon the $D_s^* \dash D_s$ mass
splitting which determines the magnitude of the neutral pion's three-momentum.
Using the improved value for this splitting recently reported by the CLEO
collaboration $M_{D_s^*}-M_{D_s} = 144.22 \pm 0.60 \MeV$ \CLEOII, we find
$| \vec{p}_{\pi^0} | = 49.0 \MeV$.  Note that we have set the parameter
$f$ in eqn.~\pizerorate\ equal to $f_\eta = 171 \MeV$ rather than
$f_\pi = 132 \MeV$ as in eqn.~\strongrate.  The difference between these two
decay constants represents an $SU(3)$ breaking effect that is higher order in
Chiral Perturbation Theory.  Our use of $f_\eta$ diminishes the magnitude of
$\Gamma(D_s^*\rightarrow D_s\pi^0)$ and provides a conservative estimate for
the impact of higher order terms in the chiral Lagrangian on the decay rate.

	It is instructive to determine the order of magnitude of the $D_s^*
\to D_s \pi^0$ branching fraction in the limit of exact $SU(3)$ symmetry.
Taking the ratio of eqn.~\pizerorate\ to \strongrate\ and using $SU(3)$ to
set $\Gamma(D_s^*\rightarrow D_s\gamma)=\Gamma(D^{*+} \rightarrow D^+ \gamma)$
and $f_\eta = f_\pi$, we find
\eqn\br{\Br(D_s^*\rightarrow D_s \pi^0) \simeq {\Gamma(D_s^* \to D_s \pi^0)
\over \Gamma(D_s^* \to D_s \gamma)} = {1\over8} \left({m_d - m_u\over m_s
 -(m_u+m_d)/2 }\right)^2 {|\vec p_{\pi^0}|^3\over |\vec
 p_{\pi^+}|^3}{\Br(D^{*+}\rightarrow D^0\pi^+)\over \Br(D^{*+}\rightarrow
 D^+\gamma)}.}
We then insert the experimentally measured branching fraction
$\Br(D^{*+}\rightarrow D^0\pi^+) = 68.1\%$ \CLEOI\ and the quark mass ratio
value from eqn.~\quarkratio\ to deduce
\eqn\isobr{\Br(D_s^*\rightarrow D_s \pi^0)\simeq
8 \times 10^{-5}/\Br(D^{*+}\rightarrow D^+ \gamma).}

	$SU(3)$ corrections to this result are likely to be significant.  In
particular, the rate for $D_s^* \to D_s \g$ is very sensitive to $SU(3)$
breaking in the transition magnetic moment $\mu_3$.  We can investigate the
general impact of $SU(3)$ violation upon radiative $D^*$ decays in the
nonrelativistic constituent quark model.    In this model, the constant $\b$
which enters into $\mu_a^{(l)}$ becomes dependent upon the subscript $a$ and
is replaced by the inverse of the constituent quark mass.  The down and
strange quark magnetic moments $\mu_2^{(l)}$ and $\mu_3^{(l)}$ are then
negative while the charm quark moment $\mu^{(h)}$ is smaller but positive.  A
partial cancellation between the heavy and light magnetic moments thus occurs
in the radiative decay $D^{*+} \to D^+ \g$ which is reflected in the small
experimental upper bound on its branching fraction
$\Br(D^{*+} \to D^+ \gamma) < 4.2 \%$ (90\% CL) \CLEOI.  The cancellation is
even stronger for $D_s^* \to D_s \gamma$.  Setting the constituent down,
strange and charm masses equal to $330 \MeV$, $550 \MeV$ and $1600 \MeV$
respectively, we obtain the constituent quark model prediction
$\mu_3 / \mu_2 = 0.32$.  The $D_s^* \to D_s \gamma$ rate therefore appears to
be quite suppressed.

	In Chiral Perturbation Theory, the leading $SU(3)$ corrections to
the transition magnetic moments are calculable and are of order $\mq^{1/2}$.
They arise from one-loop Feynman diagrams that modify the light transition
moments \Amundson:
\eqna\magmoments
$$ \eqalignno{\mu_1^{(l)} &= {2\over 3}\beta-{g^2m_K\over 4\pi f_K^2}
 -{g^2m_{\pi}\over4\pi f_{\pi}^2 } & \magmoments a \cr
\mu_2^{(l)} &= -{1\over 3}\beta+{g^2m_{\pi}\over 4\pi f_{\pi}^2}
& \magmoments b \cr
\mu_3^{(l)} &= -{1\over 3}\beta +{g^2m_K\over 4\pi f_K^2}. & \magmoments c
\cr }$$
The loop contributions to $\mu_a^{(l)}$ do not occur in the ratio $2:-1:-1$
since $m_K \ne \mpi$ and therefore violate $SU(3)$.
\foot{The difference between $f_\pi = 132 \MeV$ and $f_K = 160 \MeV$
represents a higher order $SU(3)$ breaking effect in eqn.~\magmoments{}.  We
have chosen to set the parameter $f$ equal to $f_\pi$ and $f_K$ in graphs
with pion and kaon loops respectively.}
The modified magnetic moments depend upon the two variables $\b$
and $g$ which can be related to the two independent radiative branching
fractions $\Br(D^{*0}\rightarrow D^0 \gamma)=36.4\%$ and
$\Br(D^{*+} \rightarrow D^+ \gamma) < 4.2 \%$ \CLEOI.  Once $\beta$ and $g$
are known as functions of $\Br(D^{*+} \to D^+ \g)$, we can determine the
dependence of the isospin violating branching ratio
$\Br(D_s^* \rightarrow D_s\pi^0)$ upon $\Br(D^{*+} \to D^+ \g)$ as well.
The result is plotted in \brplot.  As can be seen in the figure,
$\Br(D_s^* \to D_s \pi^0)$ generally lies in the $1 \dash 2 \%$ range.
However for $\Br(D^{*+} \to D^+ \gamma) \>\ltap\> 1 \%$, there
is a very strong cancellation between $\mu^{(h)}$ and $\mu_3^{(l)}$ which
dramatically enhances $\Br(D_s^* \to D_s \pi^0)$.  So these two
branching fractions are strongly correlated.

	$SU(3)$ corrections to the transition magnetic moments of order
$\mq$ may also be important.  Unfortunately, such corrections are not
completely calculable.  They have the general structure
\eqn\counterterm{A_a m_q \ln(m_q/\Lambda)+B_a(\Lambda) m_q }
where $\Lambda$ denotes the subtraction point.  While the coefficients $A_a$
of the chiral logarithms may readily be extracted from one-loop Feynman
diagrams \YanIII, the coefficients $B_a$ come from higher
dimension operators in the chiral Lagrangian and are unknown.  The subtraction
point dependence of the $B_a$ coefficients cancels that of the
logarithms in eqn.~\counterterm.  In ref.~\YanIII, it was noted that if the
$B_a$ terms are neglected and the subtraction point is chosen to be
approximately $1 \GeV$, then the $O(\mq)$ terms considerably enhance the
magnitude of $\mu_3^{(l)}$.  In this case, the branching ratio for
$D_s^*\rightarrow D_s \pi^{0}$ is very small.  However, the strange quark
mass is not small enough to argue that the logarithm in \counterterm\
dominates over the analytic term.  We therefore believe it is more
reasonable to regard the $O(\mq)$ terms as unknown.

	In conclusion, the branching ratio for $D_s^* \to D_s \pi^0$
that we have found lies in an experimentally interesting range.  If this
isospin violating decay is observed, its branching ratio will provide
information on the pattern of $SU(3)$ violation in radiative $D^* \to D \g$
transitions.

\bigskip\bigskip
\centerline{\bf Acknowledgments}
\bigskip

	We thank John Bartelt for a discussion which stimulated our interest
in this work.

\bigskip

\listrefs
\listfigs
\bye